\documentclass[12pt]{article}
\usepackage{epsf,epsfig}

\newcommand{\bk}{{\bf k}}
\newcommand{\bx}{{\bf x}}
\newcommand{\ee}[1]{{\rm e}^{#1}}

\title{Well-proportioned universes\\suppress CMB quadrupole}

\author{Jeffrey Weeks, Jean-Pierre Luminet,\\ Alain Riazuelo, Roland Lehoucq}

\begin{document}

\maketitle

\begin{abstract}
\noindent A widespread myth asserts that all small universe models
suppress the CMB quadrupole.  In actual fact, some models suppress
the quadrupole while others elevate it, according to whether their
low-order modes are weak or strong relative to their high-order
modes. Elementary geometrical reasoning shows that a model's
largest dimension determines the rough value $\ell_\mathrm{min}$
at which the CMB power spectrum $\ell(\ell + 1) C_\ell / 2\pi$
effectively begins;  for cosmologically relevant models,
$\ell_\mathrm{min} \leq 3$.  More surprisingly, elementary
geometrical reasoning shows that further reduction of a model's
smaller dimensions -- with its largest dimension held fixed --
serves to elevate modes in the neighborhood of $\ell_\mathrm{min}$
relative to the high-$\ell$ portion of the spectrum, rather than
suppressing them as one might naively expect. Thus among the
models whose largest dimension is comparable to or less than the
horizon diameter, the low-order $C_\ell$ tend to be relatively
weak in well-proportioned spaces (spaces whose dimensions are
approximately equal in all directions) but relatively strong in
oddly-proportioned spaces (spaces that are significantly longer in
some directions and shorter in others). We illustrate this
principle in detail for the special cases of rectangular 3-tori
and spherical spaces. We conclude that well-proportioned spaces
make the best candidates for a topological explanation of the low
CMB quadrupole observed by COBE and WMAP.
\end{abstract}

\section{Introduction}

First-year data from WMAP \cite{ben03} confirm the low CMB
quadrupole and octopole observed earlier by COBE \cite{hin96}. One
possible explanation for the striking lack of large-scale power is
that the universe might not be big enough to support
long-wavelength fluctuations \cite{ell71,ell75}.  That is, the
universe may have constant curvature but with a nontrivial global
topology and a finite volume. Over the past decade many
researchers have investigated this hypothesis, initially in the
context of flat space and a vanishing cosmological constant
\cite{ste93,oli95,lev98} but later in the more general case (for a
review see Ref.~\cite{lev02}).

A myth has arisen that non-trivial topology invariably suppresses
the quadrupole and other low-order modes. In the present article
we show this myth to be false. While some non-trivial topologies
do indeed suppress the quadrupole, others do not.  In fact some
non-trivial topologies elevate the quadrupole.  As a general
principle, among spaces whose dimensions are roughly comparable to
the horizon radius $R_{\rm LSS}$, we find that {\it
well-proportioned} spaces suppress the quadrupole (a
well-proportioned space being one whose three dimensions are of
similar magnitudes) while {\it oddly-proportioned} spaces (with
one dimension much larger or smaller than the others) elevate the
quadrupole. We illustrate this principle in the case of flat
3-tori (where roughly cubical tori suppress the quadrupole while
highly oblate or prolate tori elevate it) and the case of
spherical manifolds (where the binary polyhedral spaces suppress
the quadrupole while typical lens spaces elevate it).

\section{Mode Suppression:  Fact and Fallacy}

First the fallacy.  Start with a simply connected space $X$
(typically $X$ is the 3-sphere $S^3$, Euclidean space $E^3$ or
hyperbolic 3-space $H^3$).  Construct a closed manifold $M =
X/\Gamma$ by taking the quotient of $X$ under the action of a
group $\Gamma$. For sake of discussion assume the quotient space
$M$ is fairly small, say smaller than the last scattering surface.
Because $M$ is small, the low-order (long-wavelength) eigenmodes
of the Laplacian are suppressed.  So, according to the myth, the
CMB quadrupole (and perhaps also the octopole and other low-$\ell$
CMB multipoles) are suppressed.

Now the fact.  The low-order modes of the Laplacian are indeed
suppressed.  Where the myth goes wrong is in neglecting the fact
that high-order modes are also suppressed. According to the Weyl
asymptotic formula \cite{wey11}, for sufficiently large wave
numbers $k$, the number of modes up through $k$ is roughly
proportional to the volume of $M$. The smaller the manifold $M$,
the fewer modes it supports. The question then becomes, are the
low-order modes suppressed more or less severely than the
high-order modes? If the low-order modes are suppressed more
severely than the high-order ones, the CMB quadrupole will be
suppressed relative to the rest of the power spectrum. If the
low-order modes are suppressed less severely, the CMB quadrupole
will be elevated relative to the rest of the power spectrum.
Sections 3 and 4 will show that well-proportioned spaces suppress
the low-order modes more severely while oddly-proportioned spaces
suppress them less severely.

The reader might still feel uneasy.  After all, if passing from
$X$ to the quotient space $M$ only removes modes -- never adds
them -- how can the quadrupole possibly get elevated? The answer
is that the {\it relative} strength of the modes, not their
absolute strength, determines the shape of the power spectrum. As
an analogy, imagine reducing all prices in the world by a factor
of ten while simultaneously reducing all wages, all savings and
all debts by the same factor of ten. Obviously nothing has changed
and life proceeds as before. But if in that same scenario your own
salary is merely halved, then you are five times wealthier than
before in spite of your reduced paycheck.  The same situation
occurs with oddly-proportioned spaces:  the low-order modes get
suppressed but the high-order modes get suppressed even more, so
in a relative sense the low-order modes come out stronger.

\section{Tori}

Consider a 3-torus made from a rectangular box of width $L_x$,
length $L_y$ and height $L_z$.  An orthonormal basis for its space
of eigenmodes takes the form of planar waves
\begin{equation}
\label{PlanarWave}
 \Upsilon_\bk (\bx) = \ee{2 \pi i \bk \cdot \bx}
\end{equation}
for wave vectors
\begin{equation}
\label{RectangularTorusBasis}
 \bk = \left(\frac{n_x}{L_x}, \,
             \frac{n_y}{L_y}, \,
             \frac{n_z}{L_z} \right)
\end{equation}
where $n_x$, $n_y$ and $n_z$ are integers (see Ref.~\cite{ria03}
for full details).

\begin{figure}
\centerline{\psfig{file=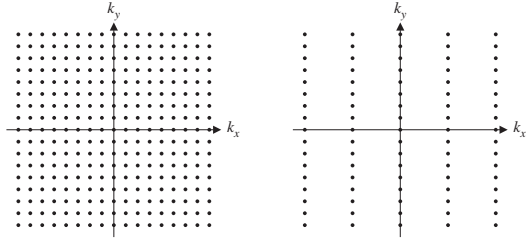,width=7cm}} \caption{In a
cubic 3-torus the allowed wavevectors $\bk$ form a cubic lattice
(left). Shrinking the torus's width by a factor $L_x/L$ stretches
the lattice in the $k_x$-direction by the inverse factor $L/L_x$
(right).  Even though the average mode density drops by a factor
of $L_x/L$, the wavevectors in the $k_y k_z$-plane remain fixed.
For ease of illustration the $k_z$-coordinate is not shown.}
\label{FigureLattice}
\end{figure}

\subsection{Cubic 3-torus}
\label{SubsectionTorusCubic}

First consider a cubic 3-torus of size $L_x$ = $L_y$ = $L_z$ =
$L$.  The allowable wave vectors (\ref{RectangularTorusBasis})
form a cubic lattice in the dual space (Fig.~\ref{FigureLattice}
left). The modulus $k = |\bk|$ of a wavevector $\bk$ is inversely
proportional to the wavelength. For the cubic 3-torus the shortest
wavevectors, corresponding to the longest wavelength, are the six
vectors $\bk = (\pm \frac{1}{L}, 0, 0)$, $(0, \pm \frac{1}{L}, 0)$
and $(0, 0, \pm \frac{1}{L})$, each of modulus $\frac{1}{L}$.  For
larger moduli $k$, corresponding to shorter wavelengths, the
number of modes of modulus $k \leq k_{\rm max}$ is simply the
number of lattice points within a ball of radius $k_{\rm max}$,
which grows in rough proportion to the ball's volume
$\frac{4\pi}{3} \, k_{\rm max}^3$.

This cubic 3-torus will serve as the standard to which other
3-tori will be compared in Sections~\ref{SubsectionTorusSmall}
through \ref{SubsectionTorusGeneral}. We assume this {\it
reference 3-torus} has a fixed size $L$ slightly larger than the
diameter of the last scattering surface.  The following sections
investigate how shrinking its dimensions suppresses its modes.
Section~\ref{SubsectionTorusSmall} shrinks all dimensions equally.
Section~\ref{SubsectionTorusOblate} shrinks one dimension while
leave the remaining two fixed.
Section~\ref{SubsectionTorusProlate} shrinks two dimensions while
leave one fixed.  Finally, Section~\ref{SubsectionTorusGeneral}
considers the general case of shrinking all dimensions, but by
different amounts.

\subsection{Small cubic 3-torus}
\label{SubsectionTorusSmall}

Start with the reference 3-torus of size $L$
(Sec.~\ref{SubsectionTorusCubic}) and shrink all its dimensions
$L_x$, $L_y$ and $L_z$ simultaneously from their original length
$L$ to some new length $L' < L$.  As the 3-torus shrinks, the
modulus of the lowest modes increases from $k = \frac{1}{L}$ to
$k' = \frac{1}{L'}$.  Indeed all modes' moduli increase by the
same factor $\frac{L}{L'}$, so the {\it relative} shape of the
mode spectrum remains unchanged, even though that spectrum now
occurs for shorter wavelengths (larger $k$).

If the original size $L$ and the final size $L'$ were both
significantly greater than the diameter of the last scattering
surface, then we would have no reason to expect any particular
effect on the CMB quadrupole.  However, in the cosmologically
interesting case that $L$ is slightly larger than the last
scattering surface while $L'$ is significantly smaller, then even
though the former lowest mode at $k = \frac{1}{L}$ would have
contributed strongly to the CMB quadrupole $C_2$, the new lowest
mode at $k' = \frac{1}{L'}$ does not, and so the quadrupole is
suppressed.  If we were to continue shrinking the 3-torus, we
would eventually lose support for the octopole $C_3$, then $C_4$
and so on.

Nevertheless, we emphasize that for the cosmologically interesting
case that $L'$ is comparable to half the diameter of the last
scattering surface, the quadrupole remains significant even though
it is suppressed relative to the reference 3-torus.  This weak but
non-trivial quadrupole will play a role in our analysis of the
general case in Section~\ref{SubsectionTorusGeneral}.

\subsection{Oblate 3-torus}
\label{SubsectionTorusOblate}

To construct an oblate 3-torus, shrink the 3-torus's width $L_x$
while leaving its length and height constant at $L_y$ = $L_z$ =
$L$.  Shrinking the 3-torus by a factor of $\frac{L_x}{L}$
stretches the lattice of allowable wavevectors
(\ref{RectangularTorusBasis}) in the dual space
(Fig.~\ref{FigureLattice} right) by the inverse ratio
$\frac{L}{L_x}$.  The lowest modes still have modulus $k =
\frac{1}{L}$ as in the reference 3-torus
(Sec.~\ref{SubsectionTorusCubic}), but their multiplicity has
dropped from 6 to 4, because now only the four wavevectors $\bk =
(0, \pm \frac{1}{L}, 0)$ and $(0, 0, \pm \frac{1}{L})$ retain that
modulus. The other two wavevectors that shared that modulus for
the cubic 3-torus, namely $(\pm \frac{1}{L}, 0, 0)$, have gotten
stretched out to $(\pm \frac{1}{L_x}, 0, 0)$ for the oblate
3-torus.

For generic modes ($k \gg \frac{1}{L_x}$), the mode density
(defined as the number of modes per unit volume in $k_x k_y
k_z$-space) drops by a factor of $\frac{L_x}{L}$.  In other words,
the mode density for the oblate 3-torus (Fig.~\ref{FigureLattice}
right) is only $\frac{L_x}{L}$ times that of the cubic reference
3-torus (Fig.~\ref{FigureLattice} left).  Visually, the mode
density corresponds to the density of the dots in
Fig.~\ref{FigureLattice}.  The crucial point here is that unlike
the lowest modes, which depend strongly on the discreteness of the
lattice, the overall strength of the high-order modes ($k \gg
\frac{1}{L_x}$) depends only on the mode density.

Comparing the results of the preceding two paragraphs, we see that
the multiplicity of the lowest-order mode $k = \frac{1}{L}$ has
dropped by a constant factor $\frac{4}{6}$ (independent of
$L_x$~!) while the overall density of modes has dropped by
$\frac{L_x}{L}$. Thus the {\it relative} strength of the lowest
mode, compared to the overall spectrum, is $\frac{4/6}{L_x/L}$ =
$\frac{2L}{3L_x}$. As $L_x$ gets small the ratio $\frac{2L}{3L_x}$
gets arbitrarily large, thus elevating the relative importance of
the lowest mode. Similar reasoning applies to the other low-order
modes.

\subsection{Prolate 3-torus}
\label{SubsectionTorusProlate}

To construct a prolate 3-torus, shrink the 3-torus's width $L_x$
and its length $L_y$ while leaving its height constant at $L_z$ =
$L$.  This stretches the lattice of allowable wavevectors
(\ref{RectangularTorusBasis}) in the dual space by a factor of
$\frac{L}{L_x}$ in the $x$-direction and a factor of
$\frac{L}{L_y}$ in the $y$-direction, for a total expansion of
$\frac{L^2}{L_x L_y}$. The lowest modes still have modulus $k =
\frac{1}{L}$ as in the reference 3-torus
(Sec.~\ref{SubsectionTorusCubic}) and the oblate 3-torus
(Sec.~\ref{SubsectionTorusOblate}), but their multiplicity has
dropped to 2, because now only the two wavevectors $\bk = (0, 0,
\pm \frac{1}{L})$ retain that modulus.

For generic modes ($k \gg \max{(\frac{1}{L_x}, \frac{1}{L_y})}$),
the mode density drops by a factor of $\frac{L_x L_y}{L^2}$. In
other words, as we pass from the cubic reference 3-torus to the
prolate 3-torus, the lattice now stretches in both the $k_x$ and
$k_y$ directions in $k_x k_y k_z$-space, and so the mode density
(the ``dot density'' in Fig.~\ref{FigureLattice}) drops by
$\frac{L_x}{L}\frac{L_y}{L} =\frac{L_x L_y}{L^2}$.

Comparing the results of the preceding two paragraphs, we see that
the multiplicity of the lowest-order mode $k = \frac{1}{L}$ has
dropped by a constant factor $\frac{2}{6}$ while the overall
density of modes has dropped by $\frac{L_x L_y}{L^2}$. Thus the
relative strength of the lowest mode, compared to the overall
spectrum, is $\frac{2/6}{L_x L_y/L^2}$ = $\frac{L^2}{3 L_x L_y}$.
As $L_x$ and $L_y$ get small the ratio $\frac{L^2}{3 L_x L_y}$
gets arbitrarily large, thus elevating the relative importance of
the low-order modes.

\subsection{Generic small 3-torus}
\label{SubsectionTorusGeneral}

Consider the case of a general rectangular 3-torus of dimensions
$L_x$, $L_y$ and $L_z$.  For cosmological interest, assume all
dimensions are comparable to or smaller than the diameter of the
last scattering surface, and in particular less than the size of
the reference 3-torus, that is $L_x$, $L_y$, $L_z < L$.

To understand this general 3-torus, imagine passing from the
reference 3-torus of size $(L, L, L)$ to the general 3-torus of
size $(L_x, L_y, L_z)$ in two steps:  first shrink isotropically
from size $(L, L, L)$ to size $(L_{\rm max}, L_{\rm max}, L_{\rm
max})$, where $L_{\rm max} = \max \, \{ L_x, L_y, L_z \}$, then
shrink anisotropically from $(L_{\rm max}, L_{\rm max}, L_{\rm
max})$ to $(L_x, L_y, L_z)$.

The first (isotropic) shrinking takes the reference 3-torus $(L,
L, L)$ to a small cubic 3-torus $(L_{\rm max}, L_{\rm max}, L_{\rm
max})$. Such shrinking suppresses the quadrupole, but if the
shrinking isn't too severe -- say $L_{\rm max}$ remains comparable
to half the horizon radius -- the quadrupole remains significant
(Section~\ref{SubsectionTorusSmall}).  The value of $L_{\rm max}$
fixes the modulus $k_{\rm min} = \frac{1}{L_{\rm max}}$ of the
torus's lowest modes, which in turn fix the value $\ell_{\rm min}$
at which the CMB power spectrum effectively begins.

The second (anisotropic) shrinking takes the small cubic 3-torus
$(L_{\rm max}$, $L_{\rm max}$, $L_{\rm max})$ to the general
3-torus $(L_x, L_y, L_z)$. The torus's largest dimension remains
fixed at $L_{\rm max}$, so the lowest mode's modulus remains at
$k_{\rm min}$, as defined in the previous paragraph, while its
multiplicity drops by at most a factor of three
(Section~\ref{SubsectionTorusProlate}). As the 3-torus's remaining
dimensions shrink, the high $k$ (short wavelength) part of the
mode spectrum is suppressed in proportion to the 3-torus's volume
(Sections~\ref{SubsectionTorusOblate} and
\ref{SubsectionTorusProlate}).  Thus whenever this anisotropic
shrinking decreases the 3-torus's volume by more than a factor of
three, we can expect the {\it relative} strength of the lowest
modes to increase relative to the high modes.  Note, though, that
the CMB power spectrum still effectively begins at the same
$\ell_{\rm min}$ as before.

In summary, the first (isotropic) shrinking pushes the entire mode
spectrum uniformly towards shorter wavelengths, moving the lowest
modes towards higher $k$.  The second (anisotropic) shrinking then
holds the (new) lowest modes fixed (with at most a factor of three
decrease in multiplicity) while it further suppresses all short
wavelength modes, in proportion to the volume of the 3-torus.  If
the first (isotropic) shrinking is mild while the second
(anisotropic) shrinking is severe (volume changes by more than a
factor of three), then the latter effect trumps the former, and
the lowest CMB multipoles ($\ell \sim \ell_{\rm min}$) are
elevated relative to their higher-$\ell$ neighbors in the
Sachs-Wolfe plateau. \\

The preceding geometrical arguments constitute a complete and
rigorous proof that the ordinary Sachs-Wolfe component of the CMB
behaves as claimed.  Nevertheless, readers may be interested to
know that careful CMB simulations, taking into account the
integrated Sachs-Wolfe and Doppler components as well as the
ordinary Sachs-Wolfe component, confirm these conclusions
\cite{ria03}.

\section{Spherical Spaces}

The 3-sphere $S^3$ supports a discrete set of modes with
eigenvalues $-k(k+2)$ and multiplicities $(k + 1)^2$ indexed by a
nonnegative integer $k$.  In terms of standard $(x, y, z, w)$
coordinates on $R^4 \supset S^3$, the modes are exactly the
homogeneous harmonic polynomials of degree $k$ in the variables
$x$, $y$, $z$ and $w$.

Taking the quotient of $S^3$ by a finite fixed point free group
$\Gamma$ of isometries yields a spherical manifold $M =
S^3/\Gamma$ of volume $\frac{{\rm Vol}(S^3)}{|\Gamma|}$.  Each
mode of $M$ lifts to a $\Gamma$-periodic mode of $S^3$, and
conversely each $\Gamma$-periodic mode of $S^3$ defines a mode of
$M$.  In practice one works with the $\Gamma$-periodic modes of
$S^3$.

The Weyl asymptotic formula asserts that the quotient manifold $M$
has, on average, $\frac{1}{|\Gamma|}$ times as many modes as
$S^3$. The question then is how do the low-order modes of $M$
compare to those of $S^3$?  Does $M$ suppress the low-order modes
more or less than the overall suppression factor of
$\frac{1}{|\Gamma|}$? The following subsections show that the
answer depends on whether the manifold is well-proportioned or
oddly-proportioned.

\subsection{Lens spaces $L(p,q)$}
\label{SubsectionSphericalLens}

The lens space $L(p,q)$ is the quotient of $S^3$ under the action
of a cyclic group $\Gamma$.  Roughly speaking, the quotient acts
in a single direction leaving the orthogonal directions
``un-quotiented''.  In this sense a lens space is analogous to the
oblate 3-torus of Sec.~\ref{SubsectionTorusOblate}.  Indeed, just
as decreasing a single dimension $L_x$ does not change the oblate
3-torus's lowest mode, increasing the order $p$ of the lens space
(thus decreasing the volume of the manifold) does not change the
lens space's lowest mode.

The 3-sphere supports a $k = 1$ mode of multiplicity four, spanned
by the four linear harmonic polynomials $\lbrace x, y, z, w
\rbrace$.  However, no linear polynomial is invariant under any
fixed point free isometry of the 3-sphere (proof:  a linear
polynomial achieves a unique maximum on $S^3$ which would need to
be preserved, contradicting the fixed point free assumption), so
no nontrivial spherical 3-manifold $S^3/\Gamma$ admits a $k = 1$
mode.  The first potentially nontrivial mode is $k = 2$.

The 3-sphere supports a $k = 2$ mode of multiplicity nine.  Within
that 9-dimensional space of modes, consider the mode given by the
quadratic harmonic polynomial
\begin{equation}
\label{LensSpacePolynomialP1}
 P_1(x,y,z,w) = x^2 + y^2 - z^2 - w^2.
\end{equation}
Geometrically $P_1$ maintains a maximum along the circle $\lbrace
x^2 + y^2 = 1, z = w = 0 \rbrace$ and a minimum along the
complementary circle $\lbrace  x = y = 0, z^2 + w^2 = 1 \rbrace$.
After a suitable change of coordinates, the generating matrix for
the holonomy of an arbitrary lens space $L(p,q)$ may be written as
\begin{equation}
\label{LensSpaceGeneratorNonhomogeneous}
  {\small
  \left(
    \begin{array}{cccc}
      \phantom{-}\cos\theta & -\sin\theta & \phantom{-}0 & \phantom{-}0  \\
      \phantom{-}\sin\theta & \phantom{-}\cos\theta & \phantom{-}0 & \phantom{-}0  \\
      \phantom{-}0 & \phantom{-}0 & \phantom{-}\cos\phi & -\sin\phi  \\
      \phantom{-}0 & \phantom{-}0 & \phantom{-}\sin\phi & \phantom{-}\cos\phi
    \end{array}
  \right)}
\end{equation}
corresponding geometrically to simultaneous rotation in the $xy$-
and $zw$-planes.  Such a rotation preserves $P_1$, so $P_1$
defines a $k = 2$ mode of {\it every} lens space $L(p,q)$, proving
that every lens space, no matter how small its volume, has a $k =
2$ mode of multiplicity at least 1.  For nonhomogeneous lens
spaces (those for which $q \neq \pm 1 \,({\rm mod} \, p)$) there
are no other invariant modes and the multiplicity is exactly 1.

For homogeneous lens spaces $L(p,\pm 1)$ the generating matrix
(\ref{LensSpaceGeneratorNonhomogeneous}) takes a special form with
$\theta = \pm\phi$ and two other modes of $S^3$ become invariant,
namely $P_2 = x z \pm y w$ and $P_3 = xw \mp yz$. (These modes
achieve their extrema along Clifford parallels, which is why they
are invariant under Clifford translations.) Thus for a homogeneous
lens space the $k = 2$ mode has multiplicity 3.

We have seen that for any lens space, homogeneous or not, the $k =
2$ mode always has multiplicity at least 1, even though the
overall mode density is suppressed in proportion to $\frac{1}{p}$.
Thus as $p$ gets large the relative importance of the $k = 2$ mode
increases.  For plausible choices of $\Omega_\mathrm{total}$ the
last scattering surface has moderate size (for example when
$\Omega_\mathrm{total} \approx 1.02$, then
$R_\mathrm{LSS}/R_\mathrm{curv} \approx 1/2$) and the lens space's
relatively strong $k = 2$ mode imprints an elevated quadrupole on
the CMB power spectrum.

Careful CMB simulations, taking into account the integrated
Sachs-Wolfe and Doppler components as well as the ordinary
Sachs-Wolfe component, confirm this expectation \cite{uza03}. The
simulations suggest that for homogeneous lens spaces $L(p,1)$ the
low-$\ell$ modes are elevated in a uniform way, while for
nonhomogeneous lens spaces $L(p,q)$, $q \neq \pm 1$ (mod $p$), the
effect on the low-$\ell$ multipoles is more erratic and depends on
the position of the observer.

\subsection{Binary polyhedral spaces}
\label{SubsectionSphericalPolyhedral}

When $\Gamma$ is the binary tetrahedral group $T^*$, the quotient
manifold $M = S^3/\Gamma$ has fundamental domain a regular
octahedron, 24 of which tile the 3-sphere.

When $\Gamma$ is the binary octahedral group $O^*$, the quotient
manifold $M = S^3/\Gamma$ has fundamental domain a truncated cube,
48 of which tile the 3-sphere.

When $\Gamma$ is the binary icosahedral group $I^*$, the quotient
manifold $M = S^3/\Gamma$ has fundamental domain a regular
dodecahedron, 120 of which tile the 3-sphere.

In all three cases the fundamental domain is well-proportioned and
we expect, in analogy with the small cubic 3-torus of
Sec.~\ref{SubsectionTorusSmall}, that the low-order modes will be
suppressed.  This expectation is borne out:  the lowest nontrivial
modes of $S^3/T^*$, $S^3/O^*$ and $S^3/I^*$ occur at $k$ = 6, 8
and 12, respectively \cite{ike95}.  In other words, as the volume
of the fundamental domain gets smaller, the lowest modes gradually
disappear.  One expects increasing suppression of the CMB
quadrupole.  A study of $S^3/I^*$ \cite{lum03} along with
work-in-progress on $S^3/T^*$ and $S^3/O^*$ confirm this
expectation.

\subsection{Spherical summary}
\label{SubsectionSphericalSummary}

Lens spaces are short in only one direction (analogous to the
oblate 3-torus) and retain a lowest mode at $k = 2$ no matter how
large the order $p$.  This effect is most pronounced for the
homogeneous lens spaces $L(p,1)$, where every observer sees his or
her own translates aligned along a single geodesic, related by
Clifford translations.  For a homogeneous lens space the lowest
mode at $k = 2$ always has multiplicity 3 (except for $L(2,1)$
where the multiplicity is 9 due to extra symmetry), and the
expected CMB quadrupole rises steadily with increasing $p$.  For a
nonhomogeneous lens space $L(p,q)$, $q \neq \pm 1$ (mod $p$), the
multiplicity of the $k = 2$ mode is only 1;  again the CMB
quadrupole is elevated but the expected effect on other low-$\ell$
modes is more erratic and depends on the observer's position in
the space.

The binary polyhedral spaces $S^3/T^*$, $S^3/O^*$ and $S^3/I^*$
are homogeneous and every observer may picture him or herself as
sitting at the centre of a well-proportioned fundamental domain
(an octahedron, truncated cube or dodecahedron, respectively).
Because the fundamental domain is well-proportioned the lowest
modes ($k < 6$, 8 or 12, respectively) disappear entirely.  For
realistic values of the cosmological parameters
($\Omega_\mathrm{total} \approx 1.02$) the Poincar\'e dodecahedral
space $S^3/I^*$ suppresses the expected quadrupole most
effectively.  To achieve comparable suppression in $S^3/T^*$ or
$S^3/O^*$ requires a slightly higher $\Omega_\mathrm{total}$.

\section{Conclusion}

The CMB quadrupole in a finite universe gets suppressed or
elevated according to whether the universe's low-order modes are
weak or strong relative to its generic high-order modes. In the
cosmologically interesting case of a fundamental domain whose
largest dimension is comparable to the diameter of the last
scattering surface, we found that in a well-proportioned universe
the low order modes will be relatively weak, while in an
oddly-proportioned universe they will be relatively strong.
Therefore well-proportioned spaces make the best candidates in the
ongoing search for a topological explanation of the low CMB
quadrupole and octopole.

\section*{Acknowledgments}

We thank the referee for help in improving the exposition.  J.W.
thanks the MacArthur Foundation for its support and thanks the
Institut d'Astrophysique de Paris and the Laboratoire de Physique
Th\'eorique at the Universit\'e Paris Sud Orsay for their
hospitality.


\label{lastpage}

\end{document}